\documentclass[epsfig,12pt]{article}
\usepackage{epsfig}
\usepackage{graphicx}
\usepackage{array}
\usepackage{color}
\usepackage{bm}
\usepackage{cite}
\usepackage{geometry}
\usepackage{latexsym}
 \usepackage{amsmath, amssymb, amscd, xypic, graphicx}

\newcommand{\beq}{\begin{equation}}   
\newcommand{\eeq}{\end{equation}}
\newcommand{\beqn}{\begin{eqnarray}}   
\newcommand{\eeqn}{\end{eqnarray}}

\newcommand{\pt}{\partial}

\def\ntwo{${\mathcal N}=2\;$}

\def\none{${\mathcal N}=1\;$}
\def\ntwot{${\mathcal N}=(2,2)\;$}
\def\ntwoo{${\mathcal N}=(0,2)\;$}

\newcommand*\xbar[1]{%
 \kern0.5ex%
  \hbox{%
   \kern0.2ex%
      \vbox{%
      \hrule height 0.5pt 
      \kern0.5ex
      \hbox{%
        \kern-0.1em
        \ensuremath{#1}%
        \kern-0.1em
      }%
    }%
  }%
}

\newcommand{\gsim}{\lower.7ex\hbox{$
\;\stackrel{\textstyle>}{\sim}\;$}}
\newcommand{\lsim}{\lower.7ex\hbox{$
\;\stackrel{\textstyle<}{\sim}\;$}}
\begin{document}

\begin{titlepage}
\begin{flushright}
FTPI-MINN-20-30, UMN-TH-3928/20
\end{flushright}

\vspace{5mm}

\begin{center}
{  \Large \bf  
Supersymmetric Proximity
}
 
 \vspace{6mm}

{\Large Mikhail Shifman} 

\vspace{3mm}

{\it  William I. Fine Theoretical Physics Institute,
University of Minnesota,
Minneapolis, MN 55455, USA}
\end{center}

\vspace{4mm}

\vspace{6mm}

\begin{center}
{\large\bf Abstract}
\end{center}

I argue that a certain perturbative proximity exists between some supersymmetric and non-supersymmetric theories (namely, pure Yang-Mills and adjoint QCD with two flavors, adjQCD$_{N_f=2}$).
I start with  ${\mathcal N}=2$ super-Yang-Mills theory built of two \none superfields: vector and chiral. In ${\mathcal N}=1$ language  the latter presents matter in the adjoint representation of SU$(N). $ Then I convert the matter superfield into a ``{\em phantom}" one (in analogy with ghosts),  breaking
${\mathcal N}=2$ down to ${\mathcal N}=1$. The global SU(2) acting between two gluinos in the original theory becomes graded. Exact results in thus deformed theory allows one to obtain insights in certain aspects of non-supersymmetric gluodynamics. In particular, it becomes clear how the splitting of the $\beta$ function
coefficients  in pure gluodynamics, $\beta_1 =(4 -\frac 13 )N$ and $\beta_2= (6-\frac 13)N^2$, occurs.  Here the first terms in the braces (4 and 6, always integers) are geometry-related while the second terms ($-\frac 13$ in both cases) are {\it bona fide} quantum effects. In the same sense adjQCD$_{N_f=2}$
is close to ${\mathcal N}=2$ SYM.

 Thus, I establish a certain proximity between  pure gluodynamics and adjQCD$_{N_f=2}$ with supersymmetric theories.  (Of course, in both cases we loose all features related to flat directions and  Higgs/Coulomb branches in ${\mathcal N}=2$.)
As a warmup exercise I use this idea in 2D CP(1) sigma model with ${\mathcal N}=(2,2)$ supersymmetry, through the minimal heterotic ${\mathcal N}=(0,2) \to$ bosonic CP(1).

\end{titlepage}

\section{Introduction}
\label{kkone}

\begin{figure}[h]
\centerline{\includegraphics[width=6cm]{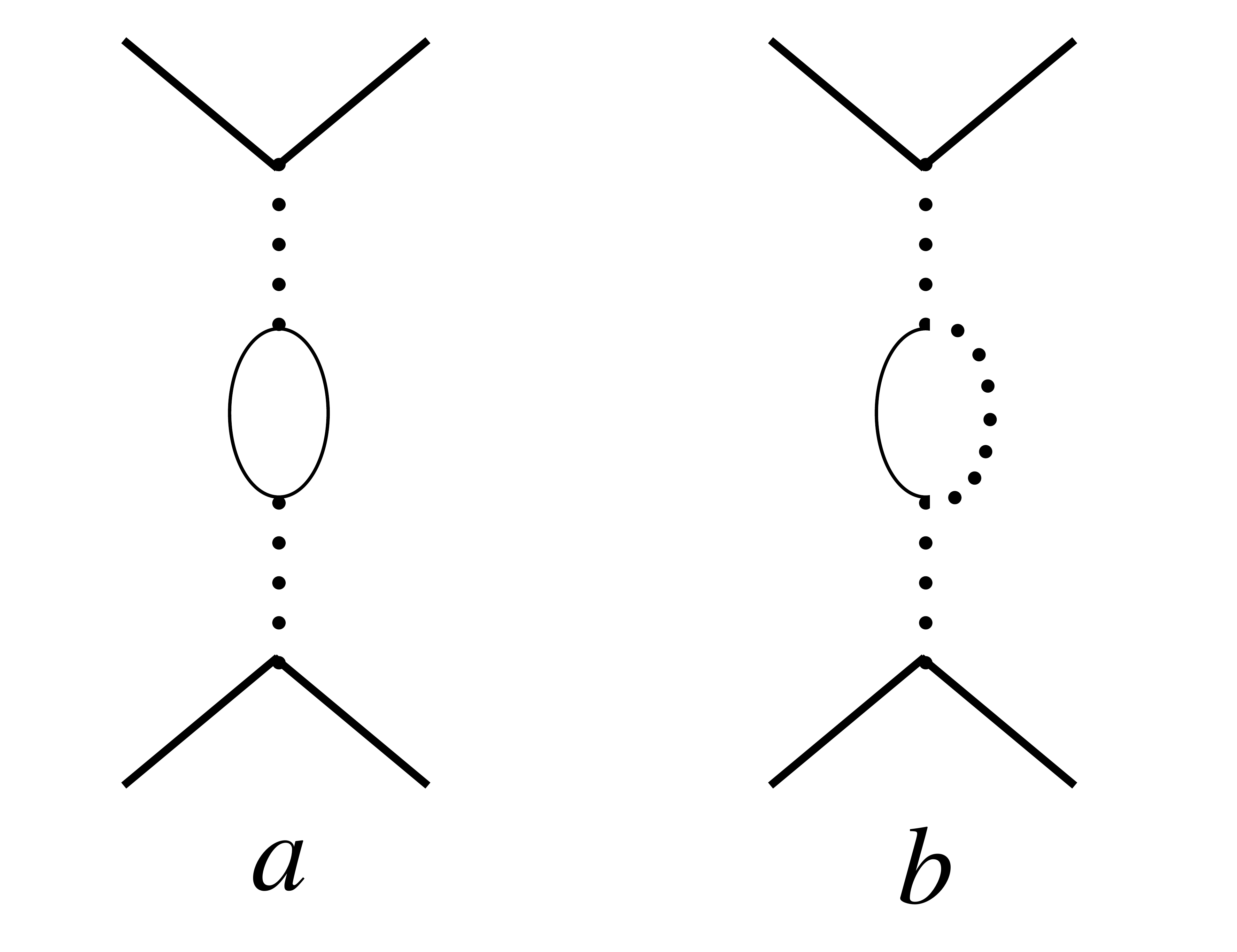}}
\caption{\small Feynman graphs for the interaction of two (infinitely) heavy probe quarks denoted by bold strait lines were calculated by Khriplovich in the Coulomb gauge. The dotted  lines stand for the (instantaneous) Coulomb interaction. Thin solid lines depict transverse gluons. In Fig. $a$ a pair of transverse gluons is produced. This graph has an
imaginary part  seen by cutting the loop. As in QED, this pair produces screening.  In Fig. $b$ similar cut in the loop is absent since it would go through a transverse gluon {\em and} the Coulomb dotted line. This graph is responsible for antiscreening.} 
\label{khri} 
\end{figure}

It has long been known that the behavior of Yang-Mills theories is unique in the sense that, unlike others, they possess asymptotic freedom. It is also known that the first coefficient of the $\beta$ function has a peculiar form,
\beq
\beta_1 = \frac{11}{3}\, N  \equiv N  \left( 4 - \frac{1}{3}\right).
\label{kone}
\eeq
where the coefficients $\beta_{1,2}$ are defined as
\beq
\beta (\alpha) =\partial_L\alpha=  -\beta_1\, \frac{\alpha^2}{2\pi} - \beta_2\, \frac{\alpha^3}{4\pi^2}+ ...\qquad \partial_L\equiv \frac{\pt}{\pt\log\mu}\,.
\label{betaf}
\eeq
The first term in the parentheses in (\ref{kone})  presents the famous {\em anti-screening} while the second is the conventional screening, as in QED. 
This was  first  noted by I.~Khriplovich who calculated \cite{khri}  the Yang-Mills coupling constant renormalization (for SU(2)) in the Coulomb gauge\,\footnote{See also 1977 papers in \cite{muz}  devoted to the same issue. Their authors apperently were unaware of Khriplovich's publication \cite{khri}.}  in 1969!
In his calculation the distinction in the origin of $4$ vs. $-\frac 13$ is transparent: the graph determining the first term in the braces does not have imaginary part, 
and hence can -- and in fact does -- produce anti-screening,
see
 (Fig. \ref{khri}).
 
 The same $4\,\,{\rm  vs.}\,  -\frac 13$ split as in (\ref{kone}) is also seen in  instanton calculus  and in calculation based on the background field method \cite{abc}. In the first case the term $4$ emerges from the zero modes  and has a geometrical meaning of the number of symmetries non-trivially realized on instanton (see below). Hence it is necessarily integer. This part of $\beta_1$  is in essence classical. {\em Bona fide} quantum corrections due to non-zero modes yield $- \frac 13$. 
 
 In the background field method 4 vs. $-\frac 13$ split in (\ref{kone}) emerges as an interplay between the magnetic (spin) vs. electric (charge) parts of the gluon vertex.
 
 Below we will discuss whether a similar interpretation exists for the second coefficient in the $\beta$ function in gluodynamics (non-supersymmetric Yang-Mills).
 To this end I will start from the simplest \ntwo super-Yang--Mills theory which  is built of two \none superfields: vector and chiral. In the  \none language  the latter presents matter in the adjoint representation of SU$(N)$. The central point is that I convert the matter superfield into a ``phantom" one (in analogy with ghosts), i.e. replace the corresponding superdeterminant by 1/superdeterminant. Then, \ntwo is broken down to \none\!\!. The global SU(2) acting between two gluinos in the original \ntwo theory becomes graded. Exact results in thus deformed theory will allow me to obtain insights in {\em non-supersymmetric gluodynamics}. 
 
\section{Setting the stage}

Supersymmetric Yang-Mills (SYM) was the first {\em four-dimensional} theory in which exact results had been obtained. In what follows I will use one of them, namely the so-called NSVZ beta function \cite {abc} in SYM theory (reviewed in \cite{Dirac}). Without matter fields we have the following result general for
\none\!\!, \ntwo and ${\mathcal N}=4$ theories:

\beq
\beta (\alpha) = - \left(n_b-\frac{n_f}{2}\right)\, \frac{ 
\alpha^2}{2\pi}\left[
1-\frac{\left(n_b-n_f\right)\,\alpha}{4\pi}\right]^{-1}\, ,
\label{totbetapg3}
\eeq
where $n_b$ and $n_f$ count  the 
gluon and gluino zero modes, respectively. For ${\mathcal N}= 1$ these numbers are
\beq
 n_b=2n_f = 4T_G =4N
 \label{k3}
 \eeq
 (I  limit myself to the SU$(N)$  gauge group theory in this note).
 For ${\cal N}= 2$,
one obtains
\beq
 n_b = n_f = 4T_G = 4N\,.
  \label{k4}
 \eeq
 Finally, for For ${\mathcal N}= 4$
 \beq
 n_b = \frac{n_f}{2} = 4T_G = 4N\,.
  \label{k4p}
 \eeq
 As it follows from (\ref{k4}) the ${\cal N}= 2$ $\beta$ function reduces to 
one-loop ($n_b=n_f$). 

The main lesson obtained in \cite{abc} was as follows. Equation (\ref{totbetapg3}) makes explicit that {\em all} coefficients
of the $\beta$ functions in {\em pure} super-Yang-Mills theories  have a geometric origin since they are in one-to-one correspondence with the number of symmetries nontrivially realized on instantons.

I will also need the extension of (\ref{totbetapg3}) including \none matter fields. We will consider one extra  \none chiral matter superfield in the {\em adjoint } representation of SU$(N)$, then
\begin{equation}
\beta (\alpha) = -\frac{\alpha^2}{2\pi}\,\,
\frac{3\,T_G -  T_G(1-\gamma)}
{1-\frac{T_G\,\alpha}{2\pi} }
\label{nsvzbetaf}
\end{equation}
where $\gamma$ is the anomalous dimension of the corresponding matter field.\footnote{For the adjoint matter superfield 
$\gamma = - \frac{N\alpha}{\pi}$.}
 This so-called NSVZ formula  appeared first in \cite{abc}, and shortly
after in a somewhat more general form in \cite{ShifmanVainshtein}.\footnote{Quite recently derivation of the NSVZ  $\beta$ function in perturbation theory has been completed in \cite{Ste}. This work also completes construction of the NSVZ scheme. It contains an extensive list of references, including those published after \cite{Step}.
}  

Needless to say, in {\em non}-supersymmetric Yang-Mills theory (gluodynamics) exact  $\beta$ function determination is  impossible. Moreover,
only the first two coefficients in the $\beta$ function are  scheme independent. In supersymmetric theories 
the all-order results mentioned above are valid in a special scheme (usually called NSVZ)  recently developed also
in perturbation theory in \cite{Ste,Step}. So far no analogue of this special scheme exists in non-supersymmetric theories. 
Therefore, in discussing below geometry-related terms in the $\beta$ function coefficients in  pure Yang-Mills theory,  I will limit myself to $\beta_1$ and$\beta_2$
which are scheme independent. One can think of extending these results to higher loops in the future.
 
\section{A simple model to begin with}
\label{kthree}

A pedagogical example of the model where the $\beta$ function similar to (\ref{nsvzbetaf}) appears is
the \ntwoo heterotic CP(1) model in two dimensions \cite{cui1}.\footnote{Of course, in CP$(N-1)$ models, even  non-supersymmetric, all coefficients have a geometric meaning, see e.g. \cite{math}. This is because such models themselves are defined through target space geometry. We will use \ntwoo heterotic CP(1) model as a toy model, a warmup before addressing Yang-Mills. Note also that minimal heterotic  models of the type (\ref{mini}), (\ref{k11})
do not exist for CP$(N-1)$ with $N>2$ because of the anomaly \cite{Moore}.} Since Ref. \cite{cui1} remains relatively unknown I will first briefly describe it in terms on \ntwoo
superfields.

The Lagrangian of the \ntwoo model in two dimentions analogous to that of \none 4D SYM is
\beq
\mathcal{L}_A = \frac 1{g^2} \int d^2 \theta_R
\frac{A^\dagger i\overset{\leftrightarrow}{\partial_{RR}} A}{1+A^\dagger A}\,, 
\label{mini}
\eeq
where $A$ is an \ntwoo bosonic chiral superfield,
\beq
A(x,\theta_R^\dagger, \theta_R) = \phi(x)+\sqrt{2}\theta_R\psi_L(x)+i\theta_R^\dagger\theta_R\partial_{LL}\phi\,.
\eeq  
Here $\phi$ is a complex scalar, and $\psi_L$ is a left-moving Weyl fermion in two dimensions.
Furthermore, the matter term is introduced through another \ntwoo superfield $B$,
\beq
B_i(x, \theta_R, \theta_R^\dagger) = \psi_{R,i}(x)+\sqrt{2}\theta_R F_i(x)+i\theta_R^\dagger\theta_R\partial_{LL}\psi_{R,i}(x)\,.
\label{defBifield}
\eeq
where $i$ is the flavor index, $i = 1,2, ..., n_f$,
and 
\beqn
\mathcal{L}_{\rm matter} &=&  \int d^2 \theta_R \,\sum_i\, \frac 12\,\frac{{B}^\dagger_i  B_i }{(1+A^\dagger A)^2}\,,
\nonumber\\[3mm]
\mathcal{L} &=& \mathcal{L}_A+\mathcal{L}_{\rm matter} \,.
\label{k11}
\eeqn
Note that the superfield $B$ contains only one physical (dynamic)  field $\psi_R$, with no bosonic counterpart. This is only possible in two dimensions.

In the minimal model (\ref{mini}) without matter the exact beta function  \cite{cui1} takes the form
\beq
\beta (g^2) = -\frac{g^4}{2\pi}\,\frac{1}{1-\frac{g^2}{4\pi}}\,,
\label{sigma3}
\eeq 
 while including matter we arrive at
 \beq
\beta (g^2) = -\frac{g^4}{2\pi}\,\frac{1+\frac{n_f\,\gamma}{2}}{1-\frac{g^2}{4\pi}}\,,
\label{sigma4}
\eeq 
 to be compared with (\ref{nsvzbetaf}). Equation (\ref{sigma3}) must be compared with (\ref{nsvzbetaf}) with the second term in the square brackets omitted.
 The parallel is apparent. 
 There are two minor but technically important distinctions, however. First, in two-dimensional  CP(1) model, unlike four-dimensional SYM,  the fermion contribution appears only in the second loop. Second, as was mentioned, the matter superfield $B$ contains only one physical degree of freedom, namely $\psi_R$. In super-Yang-Mills theory
 the matter superfield contains both components, bosonic and fermionic.
 
 Now,  if $n_f =1$ in the model under consideration supersymmetry is extended to \ntwot\!.
 In other words, in this case we deal with 
non-chiral \ntwo CP(1) which, as well-known, has only one-loop beta function. From this fact we conclude that
 \beq
 \gamma=-\frac{g^2}{2\pi}\,.
 \eeq
 Let us ask the following question: Is this information sufficient to find the first and second coefficients in non-supersymmetric (purely bosonic) CP(1) model
 without actual calculation of relevant Feynman graphs?
 
\begin{figure}[h]
\centerline{\includegraphics[width=6cm]{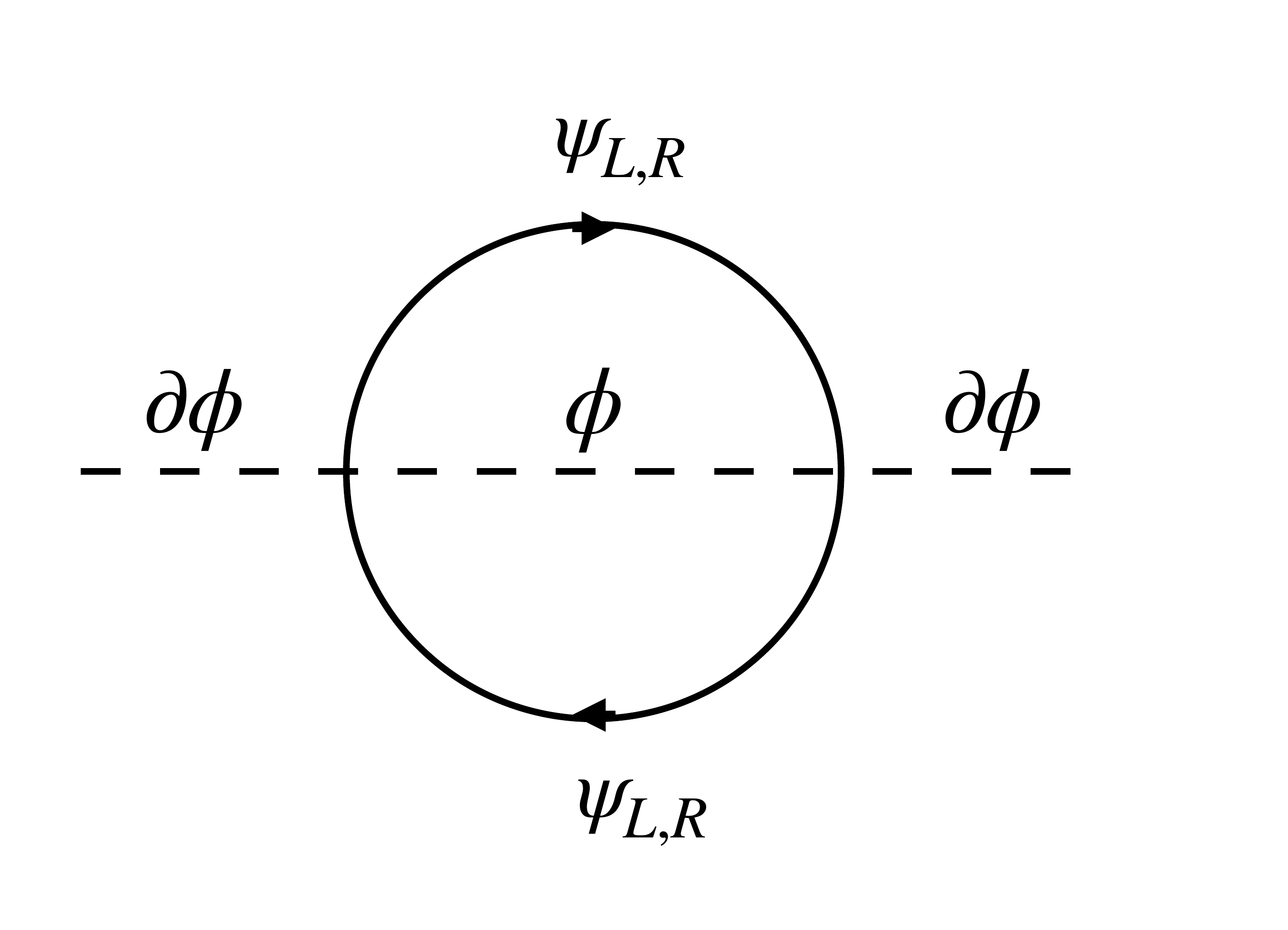}}
\caption{\small The phantom field $\psi_R$ cancels the contribution of $\psi_L$ in the two-loop beta function. This is the only diagram to be considered at small
$\phi$.}
\label{figr4} 
\end{figure}

 Surprising though it is, the answer is positive. Indeed, let us make the $B$ superfield ``phantom,"  i.e. quantize $\psi_R$ as a {\em bosonic} field. In other words, we will treat $\psi_R$ as a fermion ghost field!  In other words the extra minus sign is needed in $\psi_R$ loop.  Then the contribution of $\psi_L \in A$ and 
 $\psi_R\in B$ exactly cancel each other in the two-loop $\beta$ function (see Fig. \ref{figr4}; neither $\psi_L$ nor $\psi_R$ appear at one loop). At this stage  \ntwoo supersymmetry is preserved.
 The cancellation of $\psi_{R,L}$ contributions to $\beta$ function occurs despite the fact that one of the fermions is a left-mover and the other right-mover.
 
 Then, with the phantom $B$ superfield we  can take Eq. (\ref{sigma4}) and formally put $$n_f=-1\,.$$
 As a result we end up the with the following two-loop answer in  {\em bosonic} CP(1) model:
\beq
\beta_{\rm CP(1)\,\, bosonic} =  -\frac{g^4}{2\pi}\left(1+\frac{g^2}{2\pi}\right).
\eeq
With satisfaction we observe that the above formula coincides with the standard answer \cite{math}.
 
\section{ SU\boldmath{$(N)\,$}Yang--Mills theories} 
\label{sec4}

Now I return to gauge theories with the goal of analyzing the  second coefficient of the $\beta$ function in pure gluodynamics,
\beq
\beta_{\rm pure \,YM}  =- \frac{11}{3} \, N\,\frac{\alpha^2}{2\pi} -\frac{17}{3}\,N^2\,\frac{\alpha^3}{4\pi^2}+...
\label{k14}
\eeq
We will see that the second coefficient can be represented as follows:
\beq
\beta_2 = N^2\,\frac{17}{3}= N^2 \left(6 -\frac{1}{3}\right),
\label{k15}
\eeq
to be compared with the first coefficient in Eq. (\ref{kone}).\footnote{For the definition see Eq. (\ref{betaf}).}  The term 6 in the parentheses of (\ref{k15}) is again related to the number of instanton zero modes
in \none SYM. Thus, it has a geometrical meaning. The second term $-\frac 13$ has a {\em bona fide} quantum origin.

The line of reasoning will be the same as in Sec. \ref{kthree}. 

Let us start from \none SYM without matter. The corresponding expression is given by (\ref{nsvzbetaf}) with $\sum_i T_G(1-\gamma)=0$ or, which is the same, in 
 Eqs. (\ref{totbetapg3}) and (\ref{k3}). Then we add one chiral superfield in the adjoint representation of SU$(N)$.
By the same token as in Sec. \ref{kthree}, supersymmetry of the model with the \none adjoint matter included is automatically extended from \none  to \ntwo\!\!. In the latter the $\beta$ function is given by 
 Eqs. (\ref{totbetapg3}) and (\ref{k4}), i.e. it  reduces to one loop. 
 
The next step is  to  declare the chiral adjoint matter superfield to be ``phantom"  in \ntwo super-Yang-Mills theory. At this point  we brake the global SU(2)  acting on the doublet
 \beq
 \left(\begin{array}c
 \lambda\\
 \tilde\lambda
 \end{array}\right).
 \label{k18a}
 \eeq
 More exactly, we transform it into a graded $su(2)$ algebra  with the generators $T^\pm$ becoming odd elements, while $T^0$
 remains even. In Eq. (\ref{k18a}) $\lambda$ is the first gluino, i.e. the one from the vector superfield, while $\tilde \lambda$ is the second gluino belonging to the matter superfield.
 
Declaring the chiral superfield to be phantom  amounts to
replacing the instanton superdeterminant for \none matter by its reverse or, diagrammatically, we must change  the sign of the matter contribution in Eq. (\ref{nsvzbetaf}),  namely,
 \beq
 - T_G(1- \gamma) \to + T_G(1- \gamma)
\label{k16}
 \eeq
From the one-loop condition for \ntwo SYM $\beta$ function using Eq. (\ref{nsvzbetaf}) we derive
\beq
\gamma = -T_G\,\frac{\alpha}{\pi}\,.
\label{k20a}
\eeq
Equation (\ref{k20a}) in combination with (\ref{k16}) is to be substituted into (\ref{nsvzbetaf}). 
After this is done, the $\beta$ function in the  ``phantomized" theory takes the form (after expanding the denominator in (\ref{nsvzbetaf}))
\beqn
\beta_{\rm ph} &=& -\frac{\alpha^2}{2\pi}\, 4N\, \left(1+\frac{N\alpha}{4\pi} \right) \left(1+\frac{N\alpha}{2\pi} \right) 
= -\frac{\alpha^2}{2\pi}\, 4N\, \left(1+\frac{3N\alpha}{4\pi} \right) +...
\nonumber\\[2mm]
&\leftrightarrow & -\frac{\alpha^2}{2\pi}\, n_b\, \left[1+ \left( n_b-\frac{n_f}{2}\right)\,\frac{ \alpha}{4\pi} \right]
= -\frac{\alpha^2}{2\pi} \left[4N+ 6N^2 \,\frac{ \alpha}{2\pi} \right].
\label{k18}
\eeqn
Here $\beta$ carries a subscript ph (standing for phantom) -- it does not refer to any physical theory.
Exactly the same formula emerges from the instanton calculation in which the instanton measure is adjusted to reflect ``phantomization" of the
matter field. Since \none is unbroken, all nonzero modes still cancel. At this level the first and the second coefficients in $\beta_{\rm ph}$ are related to geometry, and are integers (see the second line above). Moreover, $n_b$ and $n_f$ in (\ref{k18}) refer to \none theory, see (\ref{k3}).

This is not the end of the story, however. Unlike the situation in Sec. \ref{kthree}, in the Yang--Mills case phantomizing the theory is not enough to pass to non-supersymmetric gluodynamics.

\begin{figure}[h]
\centerline{\includegraphics[width=9cm]{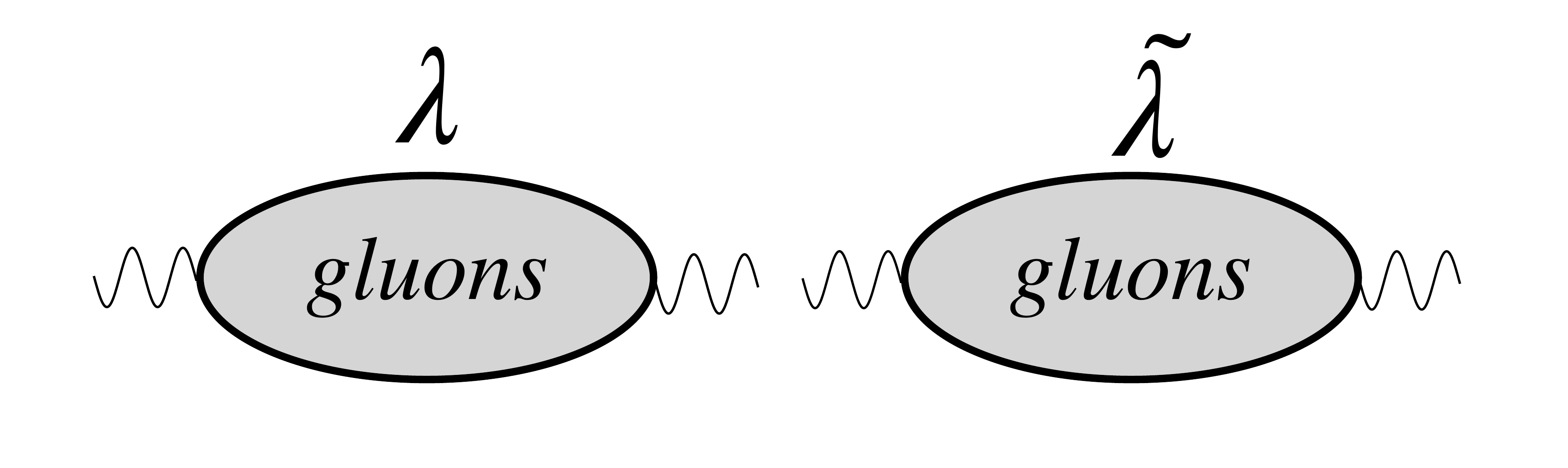}}
\caption{\small Two classes of graphs cancelling each other. Here $\lambda$ marks the gluino lines while $\tilde\lambda$ is the adjoint matter fermion from the phantom matter field. The gray shading indicates all possible gluon insertion. The wavy line is the gluon background field.}
\label{figr5} 
\end{figure}

Let us ask ourself what diagrams present in SYM but absent in non-supersymmetric gluodynamics are canceled by phantom matter? The answer is obvious and is depicted in Fig. \ref{figr5}.
Any number of gluons (with possible $\lambda\,,\tilde\lambda$ insertions) can be drawn inside the $\lambda\,,\tilde\lambda$ loops in Fig. \ref{figr5}; cancelation still persists.

\begin{figure}[h]
\centerline{\includegraphics[width=5cm]{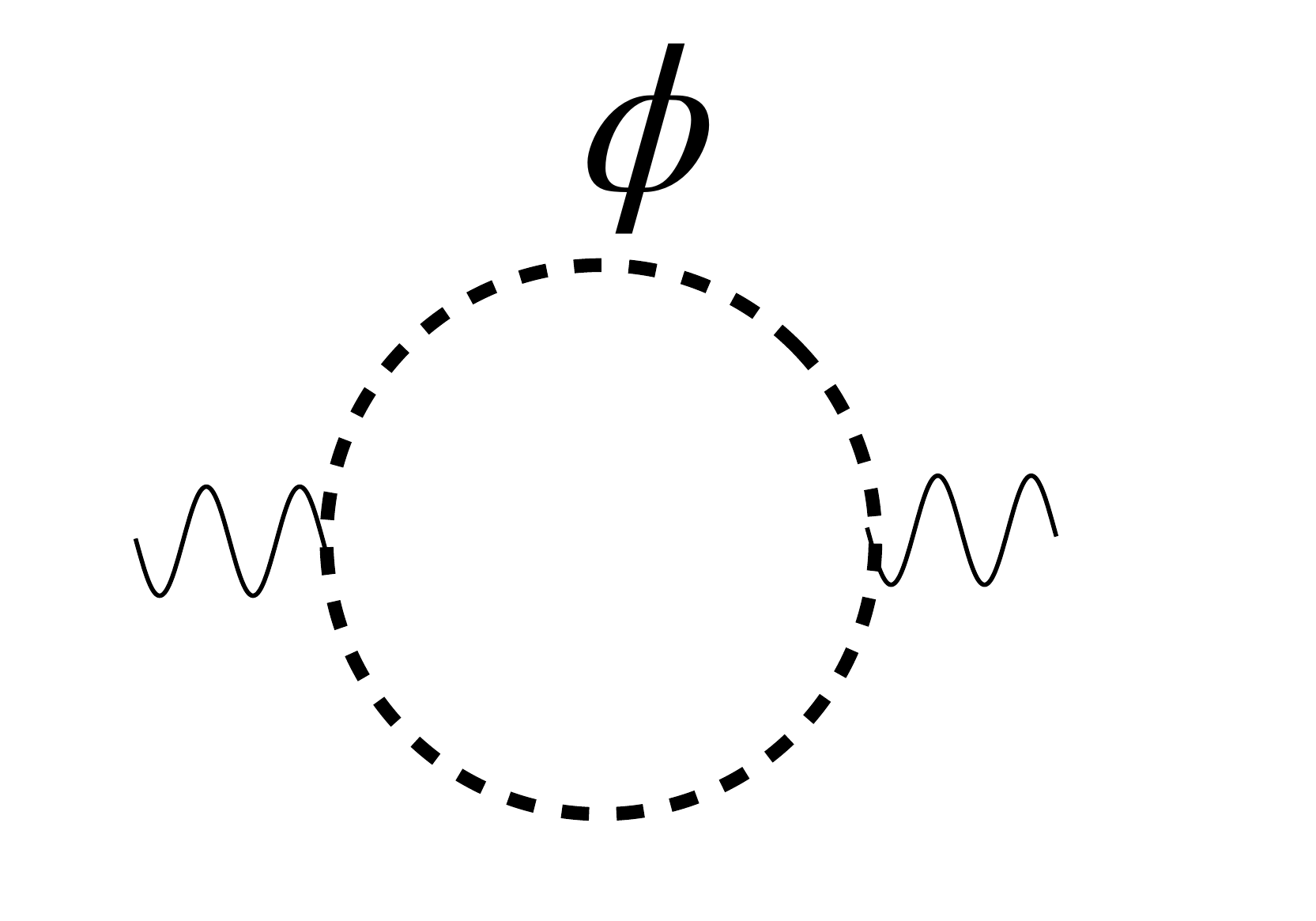}}
\caption{\small The one-loop $\phi^a$ contribution to be subtracted in passing from Eq.~(\ref{k18}) to non-SUSY Yang-Mills. }
\label{figr9} 
\end{figure}

\begin{figure}[h]
\centerline{\includegraphics[width=8cm]{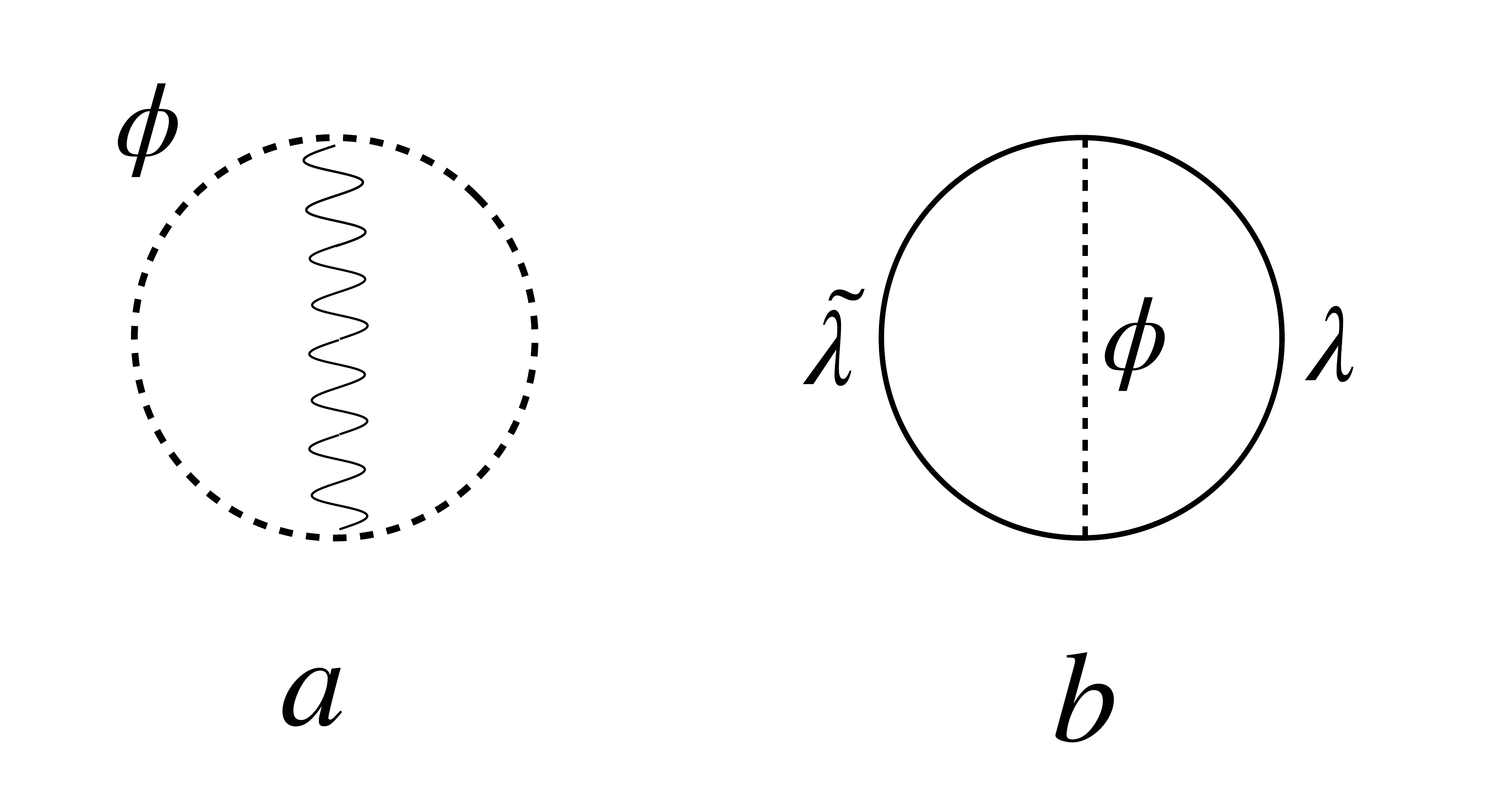}}
\caption{\small Two-loop graphs to be subtracted in passing from Eq.~(\ref{k18}) to non-SUSY Yang-Mills. 
 Here $\lambda$ marks the gluino lines while $\tilde\lambda$ is the adjoint matter fermion from the phantom superfield. The diagram $a$ combined with that in Fig. \ref{figr9} gives the $\phi^a$ contribution to two-loop $\beta$ function in QCD with scalar ``quark" (i.e. $\phi^a$ is to be considered as scalar quark in the adjoint representation.)}
\label{pt} 
\end{figure}

What does not cancel? It is obvious that all diagrams with the adjoint scalar field $\phi^a$ from the matter \none superfield still reside in Eq. (\ref{k18})
and must be subtracted in passing to non-supersymmetruc gluodynamics. The simplest example of such graph is presented in Fig. \ref{figr9}. In fact, this graph  is the only one to be dealt with at one loop.
One should not forget that in (\ref{k18}) the above diagram refers to the phantom $\phi^a$ field. Hence, its subtraction is equivalent 
to addition of the regular (unphantomized)  $\phi^a$ loop.
As for two-loop diagrams -- for brevity  I will present them omitting background field legs -- they are shown in Fig. \ref{pt}. The situation with diagram  \ref{pt}$a$  is exactly the same as with that in Fig. \ref{figr9}. The both graphs, \ref{pt}$a$ and \ref{pt}$b$  taken together, present the  phantom $\phi^a$ field contribution 
which must be subtracted from $\beta_{\rm ph}$ if we want to pass to pure gluodynamics.  In fact, because of their ``phantom" sign, to pass to pure gluodynamics, 
we must add these graphs as a normal (non-phantom)  contribution to $\beta_{\rm ph}$.

In the Appendix below I check that the impact of the quantum corrections associated with Figs.~\ref{figr9} and \ref{pt} on the phantom $\beta$ function in (\ref{k18}) is as follows:
\beq
\beta_{\rm pure \,YM} = \beta_{\rm ph}  + \delta_{\rm sc}\beta
\label{ka22}
\eeq
where
\beq
\delta_{\rm sc}\beta=
\frac 13 \left(   N\,\frac{\alpha^2}{2\pi} +N^2\,\frac{\alpha^3}{4\pi^2}\right)
\label{kar22}
\eeq
which perfectly coincides with Eqs. (\ref{kone}), (\ref{betaf}), (\ref{k14}) and (\ref{k15}). Needless to say, the sign of these corrections corresponds to screening.

\section{\boldmath{${\mathcal N}=2$} SYM and adjoint QCD with two flavors}
\label{sec5}

Recently a renewed interest in adjoint QCD led to some unexpected results (see e.g. \cite{aqcd}, and an old but useful for my present purposes review \cite{rev}).
In this section I compare \ntwo SYM with adjoint QCD with two ``quarks" (two adjoint Weyl or Majorana fields).

As well-known, in \ntwo SYM the $\beta$ function is exhausted by one loop, see Eq. (\ref{totbetapg3}) and (\ref{k4}). The first coefficient in this $\beta$ function
is 
\beq
\big(\beta_{{\mathcal N}=2}\big)_{0} = n_b -\frac 12 n_f = 2N\,.
\eeq
The geometric origin of this coefficient is obvious.

What should be changed in \ntwo SYM to convert this theory into adjoint QCD with $N_f=2$? The answer is clear -- one should subtract the same graphs in Fig. \ref{figr9}
 and \ref{pt} which were added in Eq. (\ref{ka22}),
 \beqn
 \beta_{\rm adj\, QCD} &=& \beta_{{\mathcal N}=2} - \delta_{\rm sc}\beta
 \nonumber\\[2mm]
 &=& -\left( 2+\frac{1}{3} \right)N\frac{\alpha^2}{2\pi} - \frac{N^2}{3}\,\frac{\alpha^3}{4\pi^2}\,.
 \label{kar25}
 \eeqn
Equation (\ref{kar25}) coincides with the known $\beta$ function in adjoint QCD \cite{rev}.

\section{Renormalons and adiabatic continuity}

\begin{figure}[h]
\centerline{\includegraphics[width=4cm]{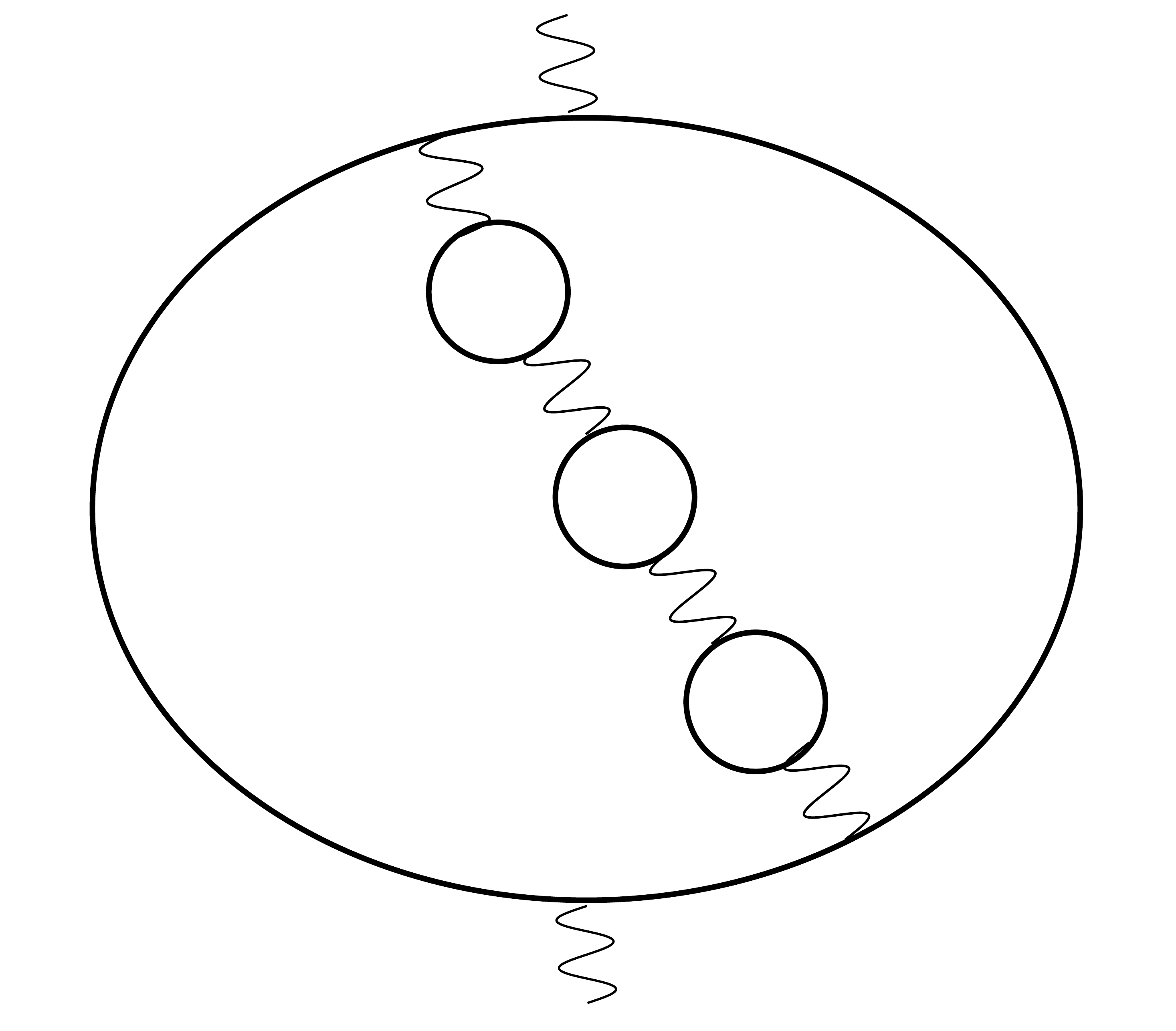}}
\caption{\small 't Hooft's bubble-chain diagrams representing renormalons.}
\label{bub} 
\end{figure}

\begin{figure}[h]
\centerline{\includegraphics[width=7cm]{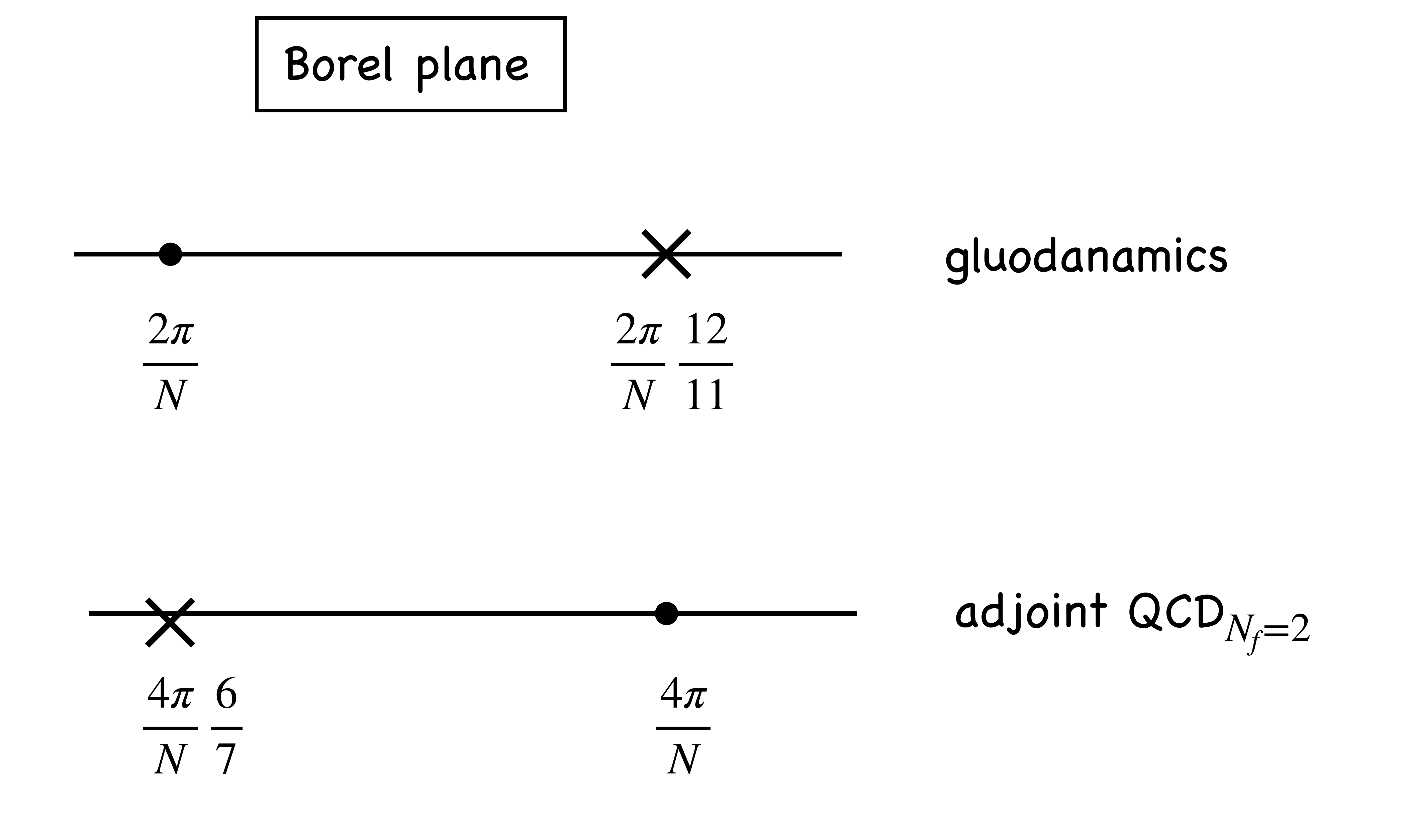}}
\caption{\small Leading renormalon-related singularities in the Borel plane (marked by crosses) for pure gluodynamics and adjoint QCD$_{N_f=2}$, respectively. 
The closed circles mark the would-be positions of the renormalon singularities if  small quantum terms $\frac 13$  in Eqs.~(\ref{kone}) and (\ref{kar25}) were ignored.}
\label{ren} 
\end{figure}

Renormalons introduced by 't Hooft \cite{renormalon} emerge from a specific narrows class of multi-loop diagrams,\footnote{For a recent brief review see \cite{cern}.}  the so-called bubble chains, see Fig. \ref{bub}.  
{\em Formally}\,\footnote{``Formally" means that in order to stay within the limits of applicability of perturbation theory we have to cut off the sum in Eq. (\ref{kar26}) at a certain value of $n=n_*$.}  this chain produces a factorially divergent perturbative series 
\beq
\sim \sum_{n} \left(\frac{\beta_0\alpha_s}{8\pi} \right)^n  n! \,.
\label{kar26}
\eeq
In the Borel plane the above factorial divergence manifests itself as a singularity at 
$$\frac{8\pi}{\beta_0} = \frac{2\pi}{N}\,\frac{12}{11}$$
in pure gluodynamics. At the same time, in adjoint QCD$_{N_f=2}$ discussed in Sec. \ref{sec5}
the renormalon-induced singularity in the Borel plane is at
$$ \frac{8\pi}{\beta_0} = \frac{4\pi}{N}\,\frac{6}{7}\,.$$
The both cases are depicted by crosses in Fig. \ref{ren}.

Simultaneously, this figure shows (by closed circles) the  would-be positions of the renormalon singularities if  small quantum terms $\frac 13$  in Eqs.~(\ref{kone}) and (\ref{kar25}) were neglected. If measured in the units of $2\pi/N$ the latter are integers. 

Why this is important? The answer to this question is associated with the program due to M. \"Unsal launched some time ago \cite{adcontinuity}
aimed at developing a quasiclassical picture on $R_3\times S_1$ at sufficiently small radius  $r(S_1)$. This program later was supplemented by the idea
{\em adiabatic continuity} stating that tending $r(S_1)\to \infty$ (i.e. returning to $R_4$) one does not encounter phase transitions on the way.
Another crucial observation 
is that the renormalon singularity must conspire with operator product expansion (OPE).  In both theories outlined 
in Fig. \ref{ren} the leading term in OPE is due to the operator $G_{\mu\nu} G^{\mu\nu}$ (the gluon condensate). 

Now, within the program \cite{adcontinuity}  a large number of new saddle-point configurations were discovered, the so-called monopole-instantons, 
also known as bions, both magnetic and neutral. Their action is $2\pi/N$, i.e. $N$ times smaller than that of instantons. 
In other words, in this picture a single instanton can be viewed as a composite state of $N$ bions.

The it becomes clear why a single bion saturates the gluon condensate in gluodynamics. Moreover, in adjoint QCD$_{N_f=2}$ similar saturation is due to bion-antibion pair which are tied up because of the existence of the fermion zero modes -- they must be contracted to give rise to  the gluon condensate. In the Borel plane (Fig. \ref{ren}) the single-bion contribution is shown by a closed circle in the upper graph, while the bion pair's contribution is depicted in the lower graph. We see that the adiabatic matching would be {\em perfect} if we could ignore the  small quantum terms $\frac 13$  in Eqs.~(\ref{kone}) and (\ref{kar25}).  The adiabatic matching is perfect in the  phantom theory of Sec. \ref{sec4}.

\section{ Conclusion and conjectures}

Starting from \ntwo super Yang-Mills and ``phantomizing" the \none matter superfield I arrived at a fully geometric $\beta_{\rm ph} $ function. This proves the 
statemnt I made in Sec. \ref{kkone} that the integer part of the first and second coefficients in $\beta_{\rm pure\,YM}$ count the number of certain symmetry generators, both bosonic and fermionic, in \none SYM. The latter theory becomes relevant because my ``phantomization" procedure breaks \ntwo $\to$ \none\!\!.  Relatively small non-integer additions represent {\em bona fide} quantum corrections which do not appear in  \none SYM because of the Bose-Fermi cancellations. At the moment one might think that this idea could shed light on some other intricate aspects of gauge theories, for instance on nuances of renormalons.

In particular, the geometric integers 4 and 6  appearing in (\ref{kone}) and (\ref{k15}) are the dimensions of the lowest-dimension gluon operators, quadratic and cubic in gluon field strength tensor, respectively. The standard renormalon wisdom says that the renormalon singularities in the Borel plane conspire
with these operators in the operator product expansion. 

At the same time the current understanding of renormalons in SYM theory continues to be incomplete (see \cite{Dunne,cern}). The gluon condensate vanishes in
supersymmetric gluodynamics. The leading renormalons have nothing to conspire with. Are there undiscovered cancellations? This suggestion does not seem likely, but we cannot avoid providing a definite answer any longer.

Summarizing, I established a certain proximity between  pure gluodynamics and an \none theory. Moreover, in the same sense {\em adjoint} QCD$_{N_f=2}$
is close to \ntwo SYM.
Conceptually this is similar to the proximity of pure gluodynamics to fundamental QCD in the limit $N\to\infty$. In the latter case the parameter governing
the proximity is adjustable, $1/N$. In the cases considered in this paper it is rather a numerical parameter whose origin is still unclear, but is quite apparent in 
Eqs. (\ref{kone}) and (\ref{k15}). It seems to be related with the dominance of magnetic interactions of gluons over their charge interactions.  If we could tend this parameter to zero we would be able to say more about the adiabatic continuity.

\vspace{2mm}

{\em Conjecture 1}

Since pure gluodynamics is close to \none phantom theory, the gluon condensate in the former must be suppressed to a certain extent since it is forbidden 
in \none supersymmetry.

\vspace{2mm}

{\em Conjecture 2}

My present consideration entangles  renormalons, their  conspiracy with  OPE, the quasiclassical treatment on $R_3\times S_1$ and adiabatic continuity all in one junction
both in pure gluodynamics and adjoint QCD$_{N_f=2}$. Can this line of studies be continued?

\vspace{2mm}

{\em Conjecture 3}

\none phantom SYM theory I have discussed in this paper calls for further investigations. For instance, I believe that
despite the presence of non-unitary contributions, if we consider amplitudes with only gluonic extermal legs their inclusive imaginary parts will 
be positive. In this narrow sense they will preserve unitarity.

\vspace{-2mm}

\section*{Acknowledgments}
I am grateful to K. Chetyrkin, A. Kataev, A. Losev, C.H. Sheu,  and K. Stepanyanz for useful communications.

This work is supported in part by DOE grant de-sc0011842.

\vspace{-1mm}

\subsection*{Appendix: Complex scalar  adjoint  quarks}

First, let us analyze diagrams in Figs. \ref{figr9} and \ref{pt}$a$. As was mentioned, they present the ``scalar quark" contributions in QCD. 
We can extract them from the know result for QCD $\beta$ function, by changing the appropriate Casimir coefficients from the fundamental representation to the 
adjoint.\footnote{I thank K. Chetyrkin, A. Kataev and K. Stepanyants for help in this point.}

The QCD $\beta$ function  with one  adjoint scalar quark extracted from \cite{KL,KL1}, in my notation reduces to 
\beq
\beta_{\rm YM + SQ}  =- \left( \frac{11}{3} \, N 
\underbrace{ -\frac{1}{3}\,N    }_{\rm Fig. 4}
\right)\frac{\alpha^2}{2\pi} 
-\left( \frac{17}{3}\,N^2  \underbrace{-2N^2 -\frac 13 N^2}_{\rm Fig.5a} \right)\frac{\alpha^3}{4\pi^2}+...
\label{kar1}
\eeq
The scalar quark contribution is underlined by underbraces.
This is not the end of the story, however. In addition, I have to take into account the graph depicted in  Fig. \ref{pt}$b$.  

The fastest and most efficient method of such calculations is the background field method, see \cite{NSVZ}.
Background field emission can occur either from the $\lambda,\,\tilde\lambda$ lines or from the $\phi$ line. In the first case the relevant propagator is 
\beq
S(x,0) =\frac{1}{2\pi^2}\,\frac{\hat{x}}{x^4}-\frac{1}{8\pi^2}\, \frac{x_\alpha}{x^2 }\, \tilde{G}_{\alpha\varphi}(0) \gamma^\varphi\gamma^5+...
\eeq
where the ellipses denote irrelevant terms in the expansion and, moreover, 
$$
\tilde{G}_{\alpha\varphi}=\frac 12\,\varepsilon_{\alpha\varphi\beta\rho}\, G^{\beta\rho}\,.
$$
In the second case the relevant propagator is that of $\phi$ (see \cite{SV}), 
\beq
G(x,0) =\frac{i}{4\pi^2}\,\frac{1}{x^2}+\frac{i}{512\pi^2}\,x^2\,G^2(0)+...
\eeq
The result of calculation of the diagram in Fig.\ref{pt}$b$ is also known in the literature. We can borrow it  from section 5 of \cite{SV},
\beq
\left. \Delta \beta \right|_{{\rm Fig. }\ref{pt}b} =  -{2N^2} \, \frac{\alpha^3}{4\pi^2}\, .
\label{k29}
\eeq
Combining (\ref{k29}) and (\ref{kar1}) we arrive at $\delta_{\rm sc}\beta$ given in Eq. (\ref{kar22}).

\end{document}